\documentclass[runningheads]{llncs}
\usepackage[T1]{fontenc}
\usepackage{graphicx}
\usepackage{amsmath}
\begin{document}
\title{Driving Computational Efficiency in Large-Scale Platforms using HPC Technologies}
\titlerunning{Driving Computational Efficiency in Large-Scale Platforms}
%
\author{A Martínez-Méndez\inst{1,2}\orcidID{0000-0002-1559-9015} \and
Antonio J. Rubio-Montero\inst{5}\orcidID{0000-0001-6497-753X} \and
Carlos J. Barrios H.\inst{1,3,4}\orcidID{0000-0002-3227-8651} \and
Hernán Asorey\inst{5}\orcidID{0000-0002-4559-8785} \and
Rafael Mayo-García\inst{5}\orcidID{0000-0002-0151-3954} \and
Luis ~A.~Núñez\inst{2}\orcidID{0000-0003-4575-5899}
}
\authorrunning{A Martínez-Méndez et al.}
%
\institute{SC3UIS-CAGE, Universidad Industrial de Santander, Bucaramanga, Colombia \and
GIRG, Universidad Industrial de Santander, Bucaramanga, Colombia \and
LIG/INRIA DataMove, Grenoble, France\and
CITI/INRIA SiNDY, Lyon, France \and
Centre for Energy, Environmental and Technological Research (CIEMAT), Madrid, Spain \\
 \email{alexander2198160@correo.uis.edu.co }, \email{cbarrios@uis.edu.co}, \email{antonio.rubio@ciemat.es}, \email{asoreyh@gmail.com}, \email{rafael.mayo@ciemat.es}, \email{lnunez@uis.edu.co}
}
\maketitle              
\begin{abstract}
The Latin American Giant Observatory (LAGO) project utilizes extensive High-Performance Computing (HPC) resources for complex astroparticle physics simulations, making resource efficiency critical for scientific productivity and sustainability. This article presents a detailed analysis focused on quantifying and improving HPC resource utilization efficiency specifically within the LAGO computational environment. The core objective is to understand how LAGO's distinct computational workloads—characterized by a prevalent coarse-grained, task-parallel execution model—consume resources in practice. To achieve this, we analyze historical job accounting data from the EGI FedCloud platform, identifying primary workload categories (Monte Carlo simulations, data processing, user analysis/testing) and evaluating their performance using key efficiency metrics (CPU utilization, walltime utilization, and I/O patterns). Our analysis reveals significant patterns, including high CPU efficiency within individual simulation tasks contrasted with the distorting impact of short test jobs on aggregate metrics. This work pinpoints specific inefficiencies and provides data-driven insights into LAGO's HPC usage. The findings directly inform recommendations for optimizing resource requests, refining workflow management strategies, and guiding future efforts to enhance computational throughput, ultimately maximizing the scientific return from LAGO's HPC investments.

\end{abstract}

%
%
%
\section{Introduction}

The Latin American Giant Observatory (LAGO) project \cite{sidelnik2017lago} represents a significant advancement in the field of astroparticle physics. It focuses on the detection of secondary particles generated by high-energy astrophysical events such as gamma-ray bursts \cite{Sidelnik2023gamma}, solar flares, and cosmic rays \cite{santos2023first}. These phenomena provide crucial insights into the universe's most energetic processes and are essential for understanding fundamental questions in physics and astrophysics. Given the complexity and scale of these research objectives, LAGO relies heavily on sophisticated simulations to model particle interactions, atmospheric effects \cite{asorey2023acorde}, and the responses of detection instruments.

The computational demands of LAGO simulations are substantial, requiring extensive processing power to analyze large datasets and perform intricate physical calculations, a common characteristic in this scientific domain as highlighted by research on high-performance computing for astrophysical simulations and astroparticle observations \cite{becerra_high-performance_2024}. As the volume of data generated grows, the efficient utilization of high-performance computing (HPC) systems becomes critical to ensure timely, accurate, and cost-effective results. These substantial computational challenges in simulation, data processing, and analysis are characteristic of modern large-scale scientific facilities, shared by experiments across domains, from particle astrophysics observatories like LAGO and the Pierre Auger Observatory \cite{allekotte2008surface} to next-generation radio astronomy projects such as the Square Kilometre Array (SKA) \cite{dewdney2009square}. Optimizing HPC resource efficiency, in addition to raw performance, is not merely a technical necessity. It is essential for enabling sustainable scientific exploration and for meeting the project’s ambitious research goals within existing computational budgets.

\begin{figure}
    \centering
    \includegraphics[width=0.45\linewidth]{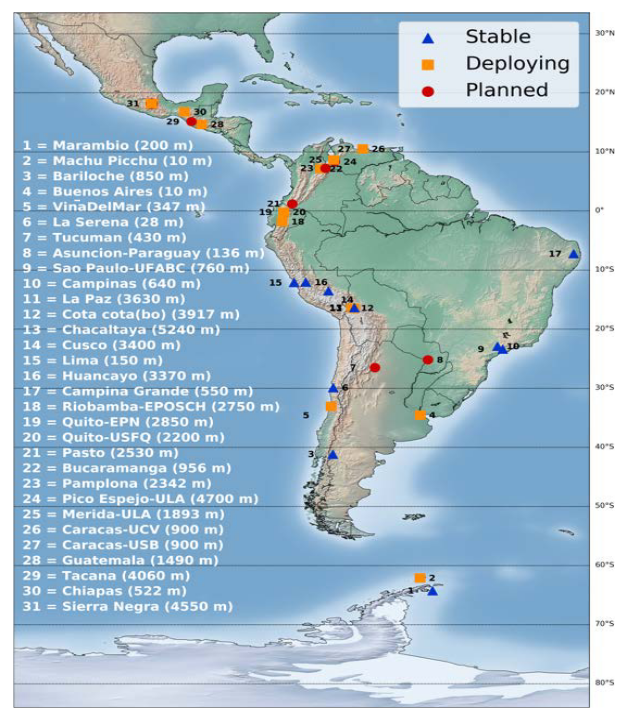}
    \caption{Map of LAGO Detector Sites across Latin America, illustrating the geographical distribution and collaborative network contributing to astroparticle physics research. Source: \cite{Rubio_Montero_2021}}
    \label{fig:lago-map}
\end{figure}

LAGO has established multiple detector sites across Latin America to effectively coordinate these efforts, each contributing valuable data to the overarching research objectives. [Fig. \ref{fig:lago-map}] shows the locations of all LAGO detector sites. This visualization highlights the project’s wide geographical reach and collaborative structure. This geographical distribution not only enhances the diversity of data collected but also underscores the need for efficient computational strategies to integrate and analyze results from these varied locations effectively.

In this context, this article emphasizes the importance of moving beyond basic performance tuning to a more comprehensive analysis of HPC resource efficiency tailored to LAGO's specific computational tasks. Understanding how different types of LAGO simulation workloads consume resources—such as CPU time, storage I/O, and network bandwidth—is fundamental. This work presents an analysis framework aimed at characterizing typical LAGO workloads and evaluating their resource utilization efficiency on HPC platforms, primarily using data extracted from execution logs and system monitoring tools. By assessing the alignment between the demands of these workloads and their actual resource consumption, we aim to identify patterns of both efficient and inefficient usage. This analysis seeks to pinpoint opportunities for optimization in areas like job configuration, resource requests, and workflow management, ultimately contributing to enhanced computational throughput and the overall sustainability of LAGO's scientific endeavors in astroparticle physics.

\section{Background}
Effective resource efficiency analysis requires understanding workload paradigms. High-Performance Computing (HPC) typically addresses tasks needing intensive, tightly-coupled parallel computation over short periods. Conversely, High-Throughput Computing (HTC) manages numerous independent, loosely-coupled tasks, often over extended durations, ideal for problems divisible into smaller units (e.g., large dataset processing or ensembles of independent simulations). While distinct, both are vital, and understanding their characteristics is key for optimizing large-scale Advanced Computing systems. The LAGO project, with its diverse computational demands, operates within this landscape.

\subsection{Brief overview of LAGO simulations and their computational complexity}

LAGO simulations cover various astroparticle phenomena (e.g., detection of gamma-ray bursts \cite{allard2008use}, space weather studies \cite{asorey2018preliminary}), demanding advanced computational methods for particle interactions, atmospheric effects, and modeling detector responses and calibration \cite{calderon2019modeling} \cite{galindo2017calibration}. High computational demands arise from processing extensive datasets and intricate physical calculations via sophisticated algorithms \cite{NunezChongo2025Convergent}. There are three main types of simulations in production, each corresponding to a specific type of data product. S0, S1, and S2. These represent, respectively: raw simulated observations in the detectors, calculated with ARTI/CORSIKA; secondary particles processed with ARTI; and simulated detector response, using Meiga/ARTI/Geant4. In addition, machine learning techniques are being developed to classify the particles that arrive at the detectors, which will constitute the future S3 data.

A cornerstone of LAGO's simulation capabilities is the ARTI framework \cite{sarmiento2022arti}, integrating tools like CORSIKA \cite{heck1998corsika}, Meiga \cite{taboada2022meiga} and Geant4 \cite{agostinelli2003geant4} for atmospheric showers and detector responses. These varying computational tasks represent distinct workloads with different resource profiles. In addition, machine learning techniques are being developed\cite{torres2024} to classify the particles that reach the detectors.

\begin{figure}
    \centering
    \includegraphics[width=0.8\linewidth]{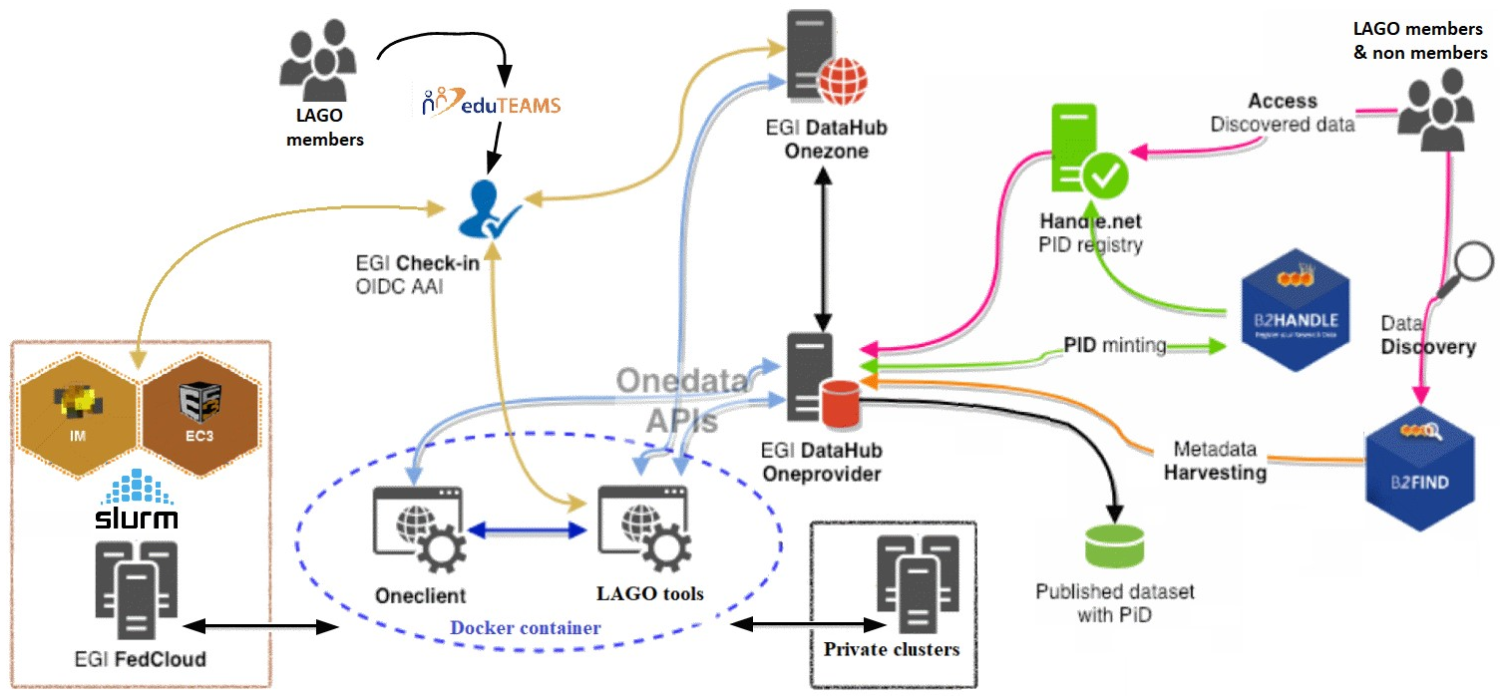} 
    \caption{Overview of the LAGO Simulation Architecture. Source: \cite{Rubio_Montero_2021}}
    \label{fig:simus-arch}
\end{figure}

LAGO's architecture (Fig.~\ref{fig:simus-arch}, source: \cite{Rubio_Montero_2021}) uses Docker images for efficiency and reproducibility \cite{Rubio_Montero_2021}. These images encapsulate core simulation functionalities, enabling automated checks, DataHub storage/publication with standardized metadata (PiDs), and B2FIND integration \cite{widmann2016eudat}. Researchers access resources via the LAGO Virtual Organization (VO) (managed by eduTEAMs \cite{coulouarn2017eduteams}). DataHub uploads are restricted to official Docker images for metadata consistency. This self-contained approach enables flexible deployment on cloud and private clusters. However, it also requires thorough efficiency analysis across different platforms. This architecture streamlines data management and collaborative research, making efficient processing key to advancing astroparticle physics \cite{NunezChongo2025Convergent}. Understanding resource efficiency within this framework is vital for maximizing its potential.

\section{Methodology}

This study employs an analysis framework to evaluate HPC resource utilization efficiency for LAGO's computational tasks, considering its prevalent coarse-grained, task-parallel execution model. The methodology relies on analyzing historical job data and system logs, interpreted with expert knowledge of LAGO workflows \cite{NunezChongo2025Convergent}, to gain insights into resource consumption patterns and identify optimization areas.

\subsection{Analysis Environment and Data Sources}\label{subsec:data_sources}
The analysis used job accounting data and system monitoring logs from the EGI FedCloud infrastructure supporting the "lagoproject.net" Virtual Organization \cite{EGIFedCloud2022}, leveraging experiences from LAGO's integration with cloud services via initiatives like EOSC-Synergy \cite{rubio2021eosc}. LAGO's significant computational activity on this infrastructure resulted in high resource consumption (e.g., Top 10 VO usage, Top 1 scratch I/O during the period) \cite{NunezChongo2025Convergent}. Key data sources included:
\begin{itemize}
    \item \textbf{Job Scheduler Logs (Slurm):} Provided data on submitted task-jobs, requested resources (CPU cores, walltime, requested memory), allocated nodes, execution time, completion status, and basic CPU usage summaries.
    \item \textbf{LAGO VO Metadata:} Information linking jobs to simulation types (e.g., CORSIKA, Geant4/Meiga), LAGO sites, workflow stages, or identifying test runs.
    \item \textbf{Expert Knowledge from the LAGO Collaboration:} Crucial for interpreting task-parallel structures, understanding workflows (e.g., those using ARTI \cite{sarmiento2022arti}), identifying site-specific variations, distinguishing test/development jobs, and validating findings \cite{NunezChongo2025Convergent}.
    \item \textbf{Filesystem Logs/Monitoring (Selective):} Used for insights into I/O patterns for representative workloads.
\end{itemize}
The software environment involved standard LAGO simulation codes in Docker containers, often managed by wrapper scripts orchestrating numerous task-jobs \cite{Rubio_Montero_2021}.

\subsection{Workload Characterization}\label{subsec:workload_characterization}
A primary step was categorizing dominant computational workloads by computation type and usage patterns, achieved by:
\begin{itemize}
    \item Analyzing job submission patterns, parameters, and LAGO metadata.
    \item Clustering jobs by resource usage profiles (CPU time per task) and duration.
    \item Explicitly differentiating production simulation campaigns (many task-jobs) from data processing and shorter 'test' or development runs using heuristics and expert input.
    \item Noting, where possible, site-specific simulation variations influencing resource needs.
\end{itemize}
This defined workload profiles (W1: 'MC Simulation Tasks', W2: 'Data Processing Tasks', W3: 'User Analysis/Development/Testing') for efficiency assessment. The workload categories previously identified were:
\begin{itemize}
    \item \textbf{W1: Large-Scale Monte Carlo (MC) simulations (e.g., CORSIKA):} Many small to medium jobs (1–4 cores), often generated via toolchains like ARTI \cite{sarmiento2022arti}. Represented $\sim$40\% of total core-hours. Efficiency stems from granular submission, avoiding idle reservations.
    \item \textbf{W2: Data Processing/Reconstruction:} Medium-duration jobs (2-12 hours), often arrays or sequential task-jobs. Varied resource requests per task, sometimes fewer cores but significant I/O. Accounted for $\sim$35\% of core-hours.
    \item \textbf{W3: User Analysis, Development, and Testing:} Heterogeneous mix including interactive jobs, analysis scripts, and frequent short "test" submissions. Variable requests, making up $\sim$25\% of core-hours but more of total job count.
\end{itemize}
Site-specific simulations add further heterogeneity.

\subsection{Resource Efficiency Metrics and Analysis}\label{subsec:metrics}
To assess workload efficiency, we focused on standard metrics per task-job, interpreted within the overall campaign/activity context:
\begin{itemize}
    \item \textbf{CPU Efficiency:} $\frac{\text{Actual CPU Time Used}}{\text{Allocated CPU Time}}$. Indicates CPU activity.
    \item \textbf{Walltime Efficiency:} $\frac{\text{Actual Walltime Used}}{\text{Requested Walltime}}$. Measures runtime estimation accuracy.
    \item \textbf{Job Completion Rate:} Percentage of successful task-jobs.
    \item \textbf{I/O Patterns (Qualitative/Quantitative):} Data volume read/written per task/campaign.
\end{itemize}
We calculated metrics for each task-job and then aggregated them by workload type. The analysis focused on metric distributions to better understand variability and the effect of outliers such as test jobs. High per-task CPU efficiency with granular submission contributes to overall campaign efficiency.

\subsection{Analysis Workflow}\label{subsec:workflow}
\begin{figure}
    \centering
    \includegraphics[width=0.5\linewidth]{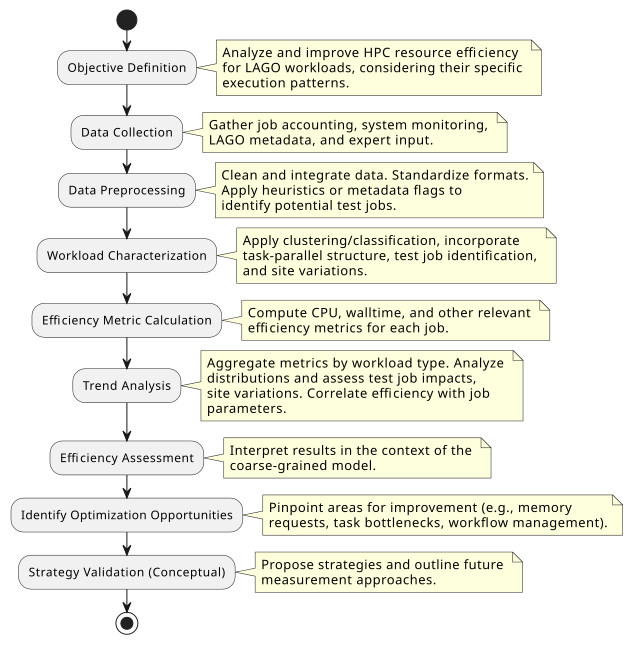}
    \caption{Conceptual Workflow for Analyzing HPC Resource Efficiency in the LAGO Framework, highlighting the steps from objective definition and data collection to workload characterization, efficiency analysis, and identification of optimization strategies.} 
    \label{fig:metod}
\end{figure}

The analysis workflow (Fig.~\ref{fig:metod}) involved:
\begin{enumerate}
    \item \textbf{Objective Definition:} Analyze and improve LAGO's HPC resource efficiency, considering its execution patterns.
    \item \textbf{Data Collection:} Gather job accounting, system monitoring, LAGO metadata, and expert input.
    \item \textbf{Data Preprocessing:} Clean, integrate data. Standardize formats. Apply heuristics/metadata flags to identify potential test jobs.
    \item \textbf{Workload Characterization:} Classify workloads considering task-parallel structure, test jobs, and site variations (see Sect.~\ref{subsec:workload_characterization}).
    \item \textbf{Efficiency Metric Calculation:} Compute per-task-job metrics.
    \item \textbf{Trend Analysis \& Efficiency Assessment:} Aggregate metrics by workload. Analyze distributions, assess impact of test jobs/site variations. Correlate efficiency with job parameters. Interpret results in the coarse-grained model context.
    \item \textbf{Identify Optimization Opportunities:} Pinpoint improvement areas (e.g., resource requests, test job practices, task-internal bottlenecks, workflow management).
    \item \textbf{Strategy Validation (Conceptual):} Propose strategies and future measurement approaches.
\end{enumerate}

\section{Resource Efficiency Analysis Results} 

This section presents HPC resource efficiency findings for LAGO tasks on the EGI FedCloud infrastructure from the beginning of simulation activities until September 2024. LAGO simulations often use a coarse-grained parallelization, decomposing into many independent task-jobs managed semi-manually, influencing efficiency metric interpretation.

Importantly, each simulation task may represent a different physical runtime, depending on the complexity of the simulated particle. Consequently, even with abundant resources, the total execution time for a simulation campaign is determined by the slowest (most complex) individual task.

Although LAGO simulations follow a standardized structure in terms of software tools and general workflow (e.g., using CORSIKA or Geant4), parameter-level variations are common across sites. These include differences in the total simulated time, the number of particles per job, or specific physics configurations. Such variability reflects the scientific goals of each site or campaign and contributes to the observed heterogeneity in workload profiles and resource usage.

Additionally, the computational cost and runtime of each simulation can depend on physical characteristics of the phenomenon and the site being modeled, particularly altitude, which influences the development of particle showers and thus affects simulation depth, granularity, and duration.

The job records also revealed the submission of test simulations, characterized by shorter runtimes and simplified configurations. While necessary for validation and debugging, these jobs tend to skew average efficiency metrics and should be considered separately in future performance analyses.

Finally, although tools like CORSIKA and Geant4 support internal parallelization (e.g., multi-threading or MPI), integration with site-specific orchestration scripts remains poorly documented. This has limited the adoption of fine-grained parallelization strategies. Resolving these issues could improve performance by enabling better intra-task parallelism and more efficient resource usage.

To illustrate the variability in execution times for tasks within a large simulation, two examples are presented showing how tasks associated with different types of particles can have significantly different execution times. These graphs highlight the impact of particle complexity on simulation times and how this can affect the total runtime of the simulation campaign.


\begin{figure}[htb]
\centering
\begin{minipage}[b]{0.45\linewidth}
    \centering
    \includegraphics[width=\linewidth]{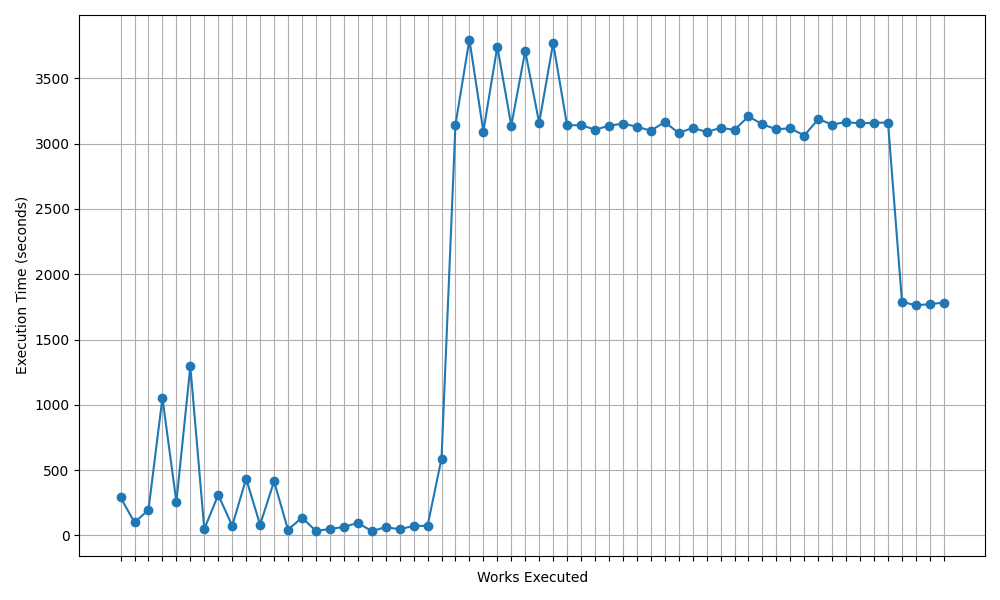}
\end{minipage}
\hspace{0.04\linewidth}
\begin{minipage}[b]{0.45\linewidth}
    \centering
    \includegraphics[width=\linewidth]{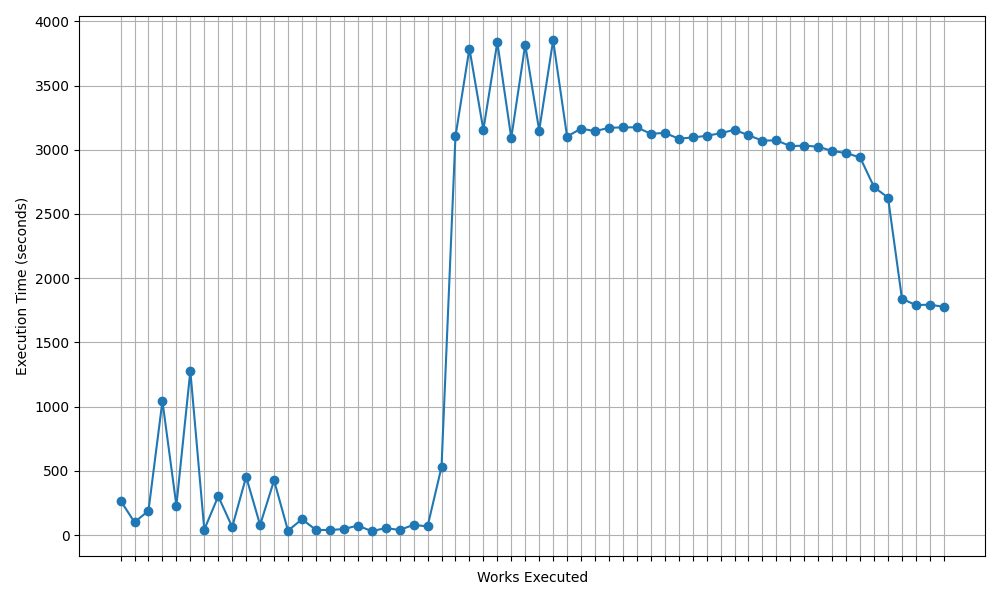}
\end{minipage}
\caption{Examples illustrating the variability in execution times per task or particle within a single simulation job. These differences reflect the complexity of individual particle simulations, which are independent and submitted as separate tasks.}
\label{fig:task-exec-differences}
\end{figure}

\subsection{Resource Utilization Patterns}

Each workload exhibited distinctive efficiency profiles:


\begin{itemize}
    \item \textbf{CPU Efficiency:} W1 (MC Simulations) tasks exhibited near-perfect CPU efficiency ($\sim$100\%), as resources are requested only when jobs are ready to run and released immediately upon completion. This on-demand strategy eliminates idle time and ensures optimal core utilization. W2 (Data Processing) jobs had moderate efficiency ($\sim$60-75\%), suggesting the presence of I/O waits or serial phases. W3 (User Analysis/Testing) showed variable and often low CPU efficiency (avg. $\sim$40\%), due to interactivity, development cycles, suboptimal scripting, and numerous test runs.
    \item \textbf{Walltime Efficiency:} W1 tasks also achieved near-perfect walltime efficiency ($\sim$100\%), as jobs are launched just-in-time and terminate as soon as computation finishes, minimizing waste. W2 and W3 jobs showed lower efficiency ($\sim$50-60\%), significantly impacted by short 'test' jobs with conservative walltime requests. Filtering out test jobs would likely improve these averages. Particle complexity also contributes to runtime spread in W1/W2.
    \item \textbf{Job Failures:} About 15\% of W1/W2 failures were due to exceeding walltime (some complex particle tasks take longer), and $\sim$10\% to out-of-memory errors (indicating issues with absolute limits or actual requirements).
\end{itemize}

Below are four examples of the temporal distribution of jobs in the simulation. These graphs demonstrate that there is no over-reservation of resources; instead, jobs only request the resources necessary for each specific task. As shown, each task has its own execution window and does not depend on the completion of other tasks to execute.

\begin{figure}[htb]
\centering

\begin{minipage}[b]{0.45\linewidth}
    \centering
    \includegraphics[width=\linewidth]{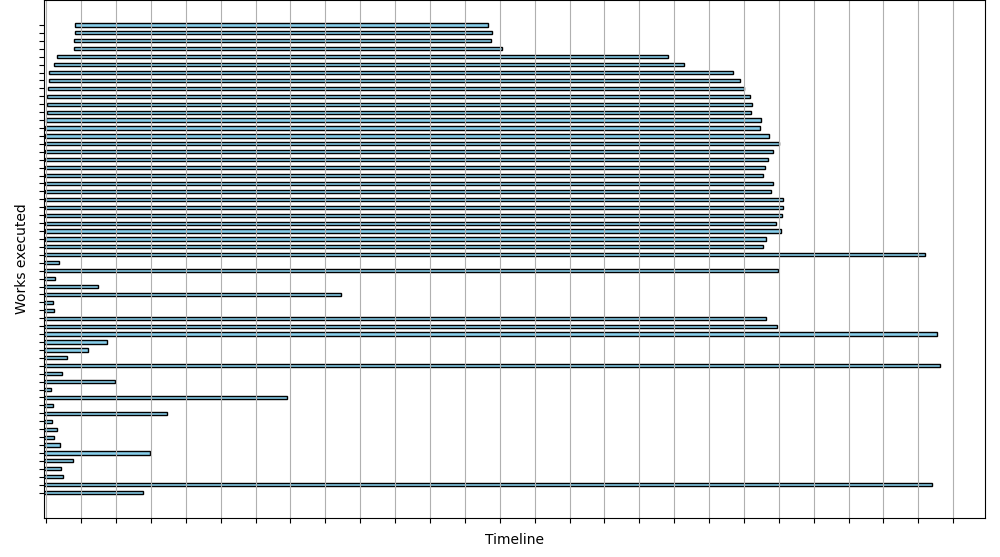}
\end{minipage}
\hspace{0.04\linewidth}
\begin{minipage}[b]{0.45\linewidth}
    \centering
    \includegraphics[width=\linewidth]{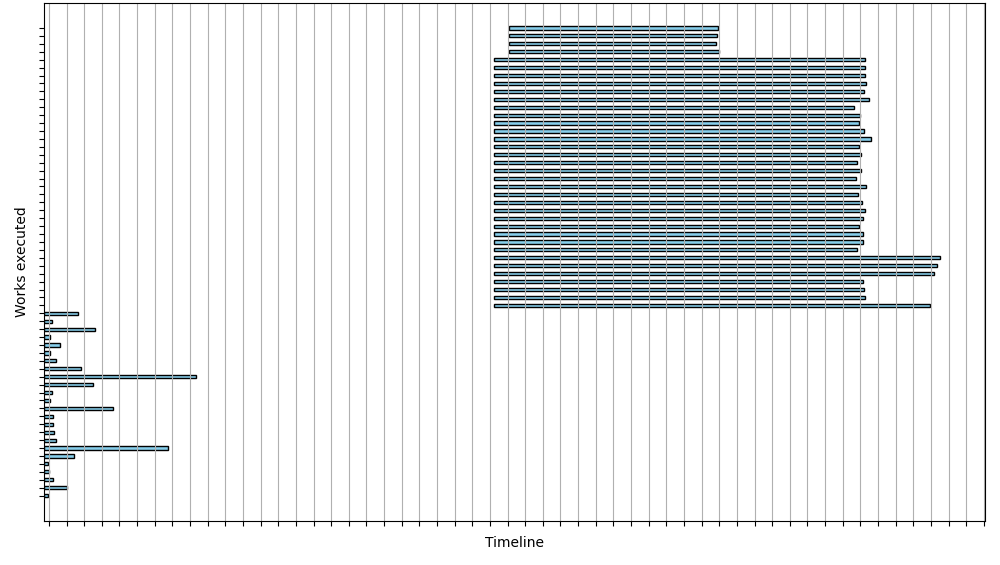}
\end{minipage}

\vspace{0.4cm}

\begin{minipage}[b]{0.45\linewidth}
    \centering
    \includegraphics[width=\linewidth]{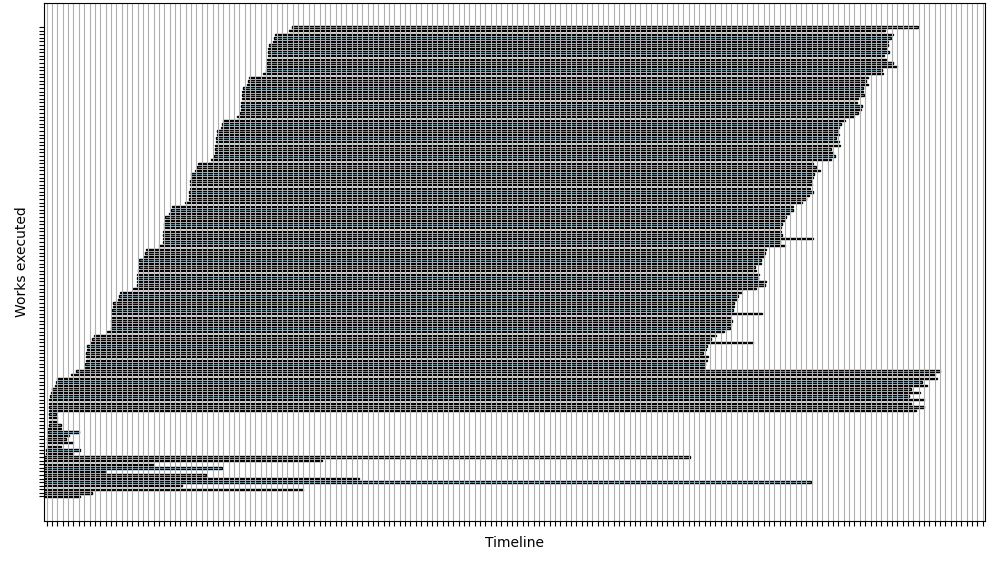}
\end{minipage}
\hspace{0.04\linewidth}
\begin{minipage}[b]{0.45\linewidth}
    \centering
    \includegraphics[width=\linewidth]{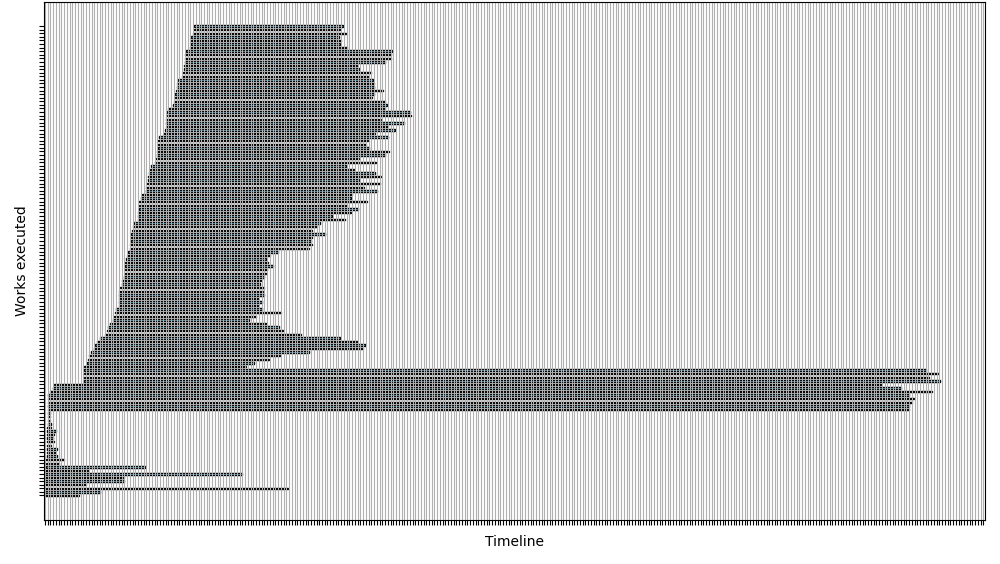}
\end{minipage}

\caption{Examples showing the distribution of simulation jobs in time. These graphs demonstrate the task-parallel execution strategy in which only the resources required by each task are requested, without idle reservations or inter-task dependencies.}
\label{fig:lago-execution-patterns}
\end{figure}

Future work should focus on (i) improving documentation and integration of fine-grained parallelization features within CORSIKA and similar tools, (ii) promoting harmonization of simulation parameters across LAGO sites where possible, and (iii) filtering test jobs in performance assessments to avoid distortions in global efficiency metrics.

\subsection{Identification of Key Inefficiencies and Challenges}
Key areas based on observed patterns and operational context include:\begin{enumerate}
    \item \textbf{Impact of Test Jobs on Averages:} Frequent short test jobs (esp. W3) skew average walltime and potentially CPU efficiency metrics, masking production run efficiency.
    \item \textbf{Suboptimal CPU Utilization in Specific Tasks/Workloads:} Moderate W2 and low W3 (excluding tests) CPU efficiency suggest tasks need optimization (I/O, serial sections), possibly compounded by difficulties leveraging internal parallelism of core simulation engines.
    \item \textbf{Heterogeneity and Lack of Standardization:} Site-specific simulation variations complicate developing universal optimization strategies.
    \item \textbf{Workflow Management Overhead:} Manual management of numerous interdependent task-jobs can add overhead and complexity, affecting time-to-solution. Campaign duration is often dictated by the longest tasks (due to particle complexity).
    \item \textbf{Altitude-Specific Simulation Needs:} Adapting simulations for LAGO's diverse detector altitudes (sea level to >5000 m.a.s.l.) alters particle interactions and computational load, impacting performance baselines across sites.
    \item \textbf{Runtime Variability due to Particle Complexity:} Within MC campaigns (W1), simulation time per primary particle varies significantly by type, energy, and shower complexity, challenging precise walltime requests.
    \item \textbf{Challenges with Internal Parallelism Integration:} Integration of internal parallelism (multi-threading/MPI in CORSIKA/Geant4) with site-specific scripts is poorly documented, limiting fine-grained parallelization and potential intra-task performance gains.
\end{enumerate}
These findings provide concrete targets for optimization efforts aimed at improving overall HPC resource utilization within the LAGO project.

\section{Discussion and Further Work}

This analysis, focusing on HPC resource efficiency for LAGO's task-parallel workloads, provides insights for improvement by moving beyond raw performance to how effectively resources are used.

\subsection{Summary of Efficiency Findings}

LAGO's task-parallel approach inherently promotes efficiency by minimizing idle resource reservations, seen in high W1 per-task CPU utilization. However, issues like variable walltime estimation accuracy persist. Test jobs significantly impact aggregate metrics, and site-specific simulation heterogeneity complicates standardized assessment. Moderate W2 CPU efficiency suggests task-internal bottlenecks.

\subsection{Implications for LAGO Computational Strategy}
The identified inefficiencies and operational characteristics have direct implications for LAGO's computational strategy:
\begin{itemize}
    \item \textbf{User Guidance/Training:} For realistic walltime requests, general resource awareness, best practices for test job submission (e.g., specific queues/annotations), and understanding task duration variance.
    \item \textbf{Targeted Profiling/Optimization:} For W2 tasks, focusing on I/O or serial sections, and understanding task complexity variance.
    \item \textbf{Resource Policy Adjustments:} Policies encouraging accurate requests; queue structures better handling many short/medium task-jobs.
    \item \textbf{Workflow Management Tools:} Adopting sophisticated tools to automate task-job management and dependency tracking.
    \item \textbf{Standardization Efforts:} Encouraging standardized setups for common tasks across sites where feasible.
\end{itemize}
Addressing these points can lead to more effective use of allocated time and resources, enhancing LAGO's scientific output.

\subsection{Challenges and Limitations}
This analysis faces challenges inherent in interpreting system-level logs for task-parallel workflows:
\begin{itemize}
    \item \textbf{System-Level View Limitations:} Scheduler logs offer a limited view (e.g., peak memory accuracy, CPU time attribution in parallel jobs).
    \item \textbf{Attributing Campaign Efficiency:} To assess the efficiency of an entire campaign, both per-task efficiency and the effectiveness of task-parallel decomposition must be considered. However, job metrics capture only part of this picture.
    \item \textbf{Impact of Heterogeneity and Test Jobs:} Site-specific simulations and test jobs complicate average efficiency calculations and global recommendations.
    \item \textbf{Manual Workflow Aspects:} Manual work distribution hinders automatic capture of full campaign structure from logs.
    \item \textbf{Challenges with Internal Parallelism Integration:} While core simulation tools like Geant4 and CORSIKA offer internal parallelism options (e.g., multi-threading) , their integration with site-specific orchestration scripts remains poorly documented within the LAGO framework. This has limited the adoption of fine-grained parallelization strategies , due to potential conflicts with existing custom scripts and the established coarse-grained, multi-job execution model. This limitation hinders potential intra-task performance gains.
\end{itemize}

\subsection{Future Directions for Efficiency Improvement}
Future work should focus on overcoming these limitations and proactively improving efficiency within LAGO's operational model:
\begin{itemize}
    \item \textbf{Enable Finer-Grained Parallelism:} Resolve conflicts to document and enable internal parallelism (e.g., multi-threading in Geant4) in core simulation engines, complementing the coarse-grained approach.
    \item \textbf{Develop Resource Estimation Tools:} For users to better estimate runtime and other key resource needs.
    \item \textbf{Integrate Application-Level Monitoring:} Using lightweight profiling/monitoring in LAGO Docker containers for granular application behavior data.
    \item \textbf{Enhanced Logging/Metadata:} Methods for users to tag "test" vs. "production" jobs and link task-jobs to campaigns/sites.
    \item \textbf{Workflow Analysis Tools:} For multi-job, task-parallel workflows using scheduler data and enhanced metadata.
    \item \textbf{Promote Workflow Management Systems:} Evaluate and promote adoption of systems like Snakemake, Nextflow for LAGO.
    \item \textbf{Explore Container/Task Optimizations:} Reduce startup times or overheads of many containerized task-jobs.
\end{itemize}

In conclusion, systematically analyzing HPC resource efficiency, while adapting interpretation to LAGO's specific task-parallel strategy, yields crucial insights. Understanding how resources are used—beyond just raw metrics—makes it possible to apply targeted optimizations. This improves throughput, supports sustainable resource use, and ultimately accelerates scientific discovery. Implementing strategies from this analysis is vital for maximizing LAGO's HPC return on investment.

\begin{credits}
\subsubsection{\ackname} The authors thank the CYTED co-funded LAGO-INDICA network (524RT0159-LAGO-INDICA: Infraestructura digital de ciencia abierta) and the EL-BONGO physics project for their support and collaboration initiatives. We also acknowledge the Universidad Industrial de Santander (UIS), especially the research groups Cómputo Avanzado y a Gran Escala (CAGE) and Grupo de Investigación en Relatividad y Gravitación (GIRG), for their institutional support and resources provided for this work. AI technology was used to proofread and polish this manuscript.(OpenAI, 2025. Gemini 2025).

\end{credits}

\bibliographystyle{IEEEtran} 
\bibliography{references}    

@inproceedings{Rubio_Montero_2021,
   title={A Novel Cloud-Based Framework For Standardized Simulations In The Latin American Giant Observatory (LAGO)},
   DOI={10.1109/wsc52266.2021.9715360},
   booktitle={2021 Winter Simulation Conference (WSC)},
   publisher={IEEE},
   author={Rubio-Montero, Antonio Juan and Pagan-Munoz, Raul and Mayo-Garcia, Rafael and Pardo-Diaz, Alfonso and Sidelnik, Ivan and Asorey, Hernan},
   year={2021},
   month=dec 
}

@article{agostinelli2003geant4,
  title={GEANT4—a simulation toolkit},
  author={Agostinelli, Sea and Allison, John and Amako, K al and Apostolakis, John and Araujo, Henrique and Arce, Pedro and Asai, Makoto and Axen, D and Banerjee, Swagato and Barrand, GJNI and others},
  journal={Nuclear instruments and methods in physics research section A: Accelerators, Spectrometers, Detectors and Associated Equipment},
  volume={506},
  number={3},
  pages={250--303},
  year={2003},
  publisher={Elsevier}
}

@article{allard2008use,
  title={Use of water-Cherenkov detectors to detect gamma ray bursts at the Large Aperture GRB Observatory (LAGO)},
  author={Allard, D and Allekotte, Ingomar and Alvarez, C and Asorey, H and Barros, H and Bertou, X and Burgoa, O and Berisso, M Gomez and Mart{\'\i}nez, O and Loza, P Miranda and others},
  journal={Nuclear Instruments and Methods in Physics Research Section A: Accelerators, Spectrometers, Detectors and Associated Equipment},
  volume={595},
  number={1},
  pages={70--72},
  year={2008},
  publisher={Elsevier}
}

@article{galindo2017calibration,
  title={Calibration of a large water-Cherenkov detector at the Sierra Negra site of LAGO},
  author={Galindo, Aline and Moreno, E and Carrasco, E and Torres, I and Carrami{\~n}ana, A and Bonilla, M and Salazar, H and Conde, R and Alvarez, W and Alvarez, C and others},
  journal={Nuclear Instruments and Methods in Physics Research Section A: Accelerators, Spectrometers, Detectors and Associated Equipment},
  volume={861},
  pages={28--37},
  year={2017},
  publisher={Elsevier}
}

@article{calderon2019modeling,
  title={Modeling the LAGO’s detectors response to secondary particles at ground level from the Antarctic to Mexico},
  author={Calder{\'o}n-Ardila, R and Jaimes-Motta, A and Sarmiento-Cano, C and Su{\'a}rez-Dur{\'a}n, M and V{\'a}squez-Ram{\'\i}rez, A and others},
  journal={36th ICRC},
  year={2019}
}

@article{heck1998corsika,
  title={CORSIKA: A Monte Carlo code to simulate extensive air showers},
  author={Heck, Dieter and Knapp, Johannes and Capdevielle, JN and Schatz, G and Thouw, T and others},
  year={1998},
  publisher={Forschungszentrum Karlsruhe Karlsruhe}
}

@article{NunezChongo2025Convergent,
  author    = {N\'{u}{\~n}ez-Chongo, Osiris and Asorey, Hern\'{a}n and Rubio-Montero, Antonio Juan and Su\'{a}rez-Dur\'{a}n, Mauricio and Mayo-Garc\'{i}a, Rafael and Carretero, Manuel},
  title     = {Convergent data-driven workflows for open radiation calculations: an exportable methodology to any field},
  journal   = {The Journal of Supercomputing},
  year      = {2025},
  volume    = {81},
  pages     = {465},
  doi       = {10.1007/s11227-024-06894-0},
  publisher = {Springer Science+Business Media, LLC, part of Springer Nature}
}

@misc{EGIFedCloud2022,
  author       = {{European Grid Infrastructure}},
  title        = {EGI FedCloud},
  year         = {2022},
  url          = {https://www.egi.eu/egi-infrastructure/},
  note         = {Accessed: 2025-04-30}
}

@article{dewdney2009square,
  title={The square kilometre array},
  author={Dewdney, Peter E and Hall, Peter J and Schilizzi, Richard T and Lazio, T Joseph LW},
  journal={Proceedings of the IEEE},
  volume={97},
  number={8},
  pages={1482--1496},
  year={2009},
  publisher={IEEE}
}

@article{allekotte2008surface,
  title={The surface detector system of the Pierre Auger Observatory},
  author={Allekotte, Ingo and Barbosa, AF and Bauleo, P and Bonifazi, C and Civit, B and Escobar, CO and Garc{\'\i}a, B and Guedes, G and Berisso, M G{\'o}mez and Harton, JL and others},
  journal={Nuclear Instruments and Methods in Physics Research Section A: Accelerators, Spectrometers, Detectors and Associated Equipment},
  volume={586},
  number={3},
  pages={409--420},
  year={2008},
  publisher={Elsevier}
}

@inproceedings{widmann2016eudat,
  title={EUDAT B2FIND: a cross-discipline metadata service and discovery portal},
  author={Widmann, Heinrich and Thiemann, Hannes},
  booktitle={EGU General Assembly Conference Abstracts},
  pages={EPSC2016--8562},
  year={2016}
}

@inproceedings{coulouarn2017eduteams,
  title={eduTEAMS un service G{\'E}ANT pour les {\'e}quipes virtuelles},
  author={Coulouarn, Tangui},
  booktitle={JRES (Journ{\'e}es r{\'e}seaux de l'enseignement et de la recherche) 2017},
  year={2017}
}

@incollection{becerra_high-performance_2024,
	address = {Cham},
	title = {High-{Performance} {Computing} for {Astrophysical} {Simulations} and {Astroparticle} {Observations}},
	volume = {1887},
	isbn = {978-3-031-52185-0 978-3-031-52186-7},
	language = {en},
	urldate = {2025-05-15},
	booktitle = {High {Performance} {Computing}},
	publisher = {Springer Nature Switzerland},
	author = {Becerra, L. M. and Sarmiento-Cano, C. and Martínez-Méndez, A. and Dominguez, Y. and Núñez, L. A.},
	editor = {Barrios H., Carlos J. and Rizzi, Silvio and Meneses, Esteban and Mocskos, Esteban and Monsalve Diaz, Jose M. and Montoya, Javier},
	year = {2024},
	doi = {10.1007/978-3-031-52186-7_13},
	note = {Series Title: Communications in Computer and Information Science},
	pages = {184--196},
}

@inproceedings{rubio2021eosc,
      address = {Berlin, Germany},
      author         = {Rubio-Montero, A.J and Pag\'an-Mu\~noz, R. and others},
      title          = "{The EOSC-Synergy cloud services implementation for the Latin American Giant Observatory (LAGO)}",
      booktitle      = "37th ICRC",
      collaboration  = "LAGO",
      volume         =  395,          
      pages          = "PoS(ICRC2021)261",
      year           = "2021",
      doi            = "10.22323/1.395.0261"
}

@article{sidelnik2017lago,
  title={{LAGO: The Latin American giant observatory}},
  author={Sidelnik, Iv{\'a}n and Asorey, Hern{\'a}n and Lago Collaboration and others},
  journal={Nuclear Instruments and Methods in Physics Research Section A: Accelerators, Spectrometers, Detectors and Associated Equipment},
  volume={876},
  pages={173--175},
  year={2017},
  publisher={Elsevier},
  doi = {10.1016/j.nima.2017.02.069}
}

@article{asorey2018preliminary,
  author = {Asorey, H. and N\'u\~nez, L. A. and Su\'arez-Dur\'an, M.},
  title = {{Preliminary Results From the Latin American Giant Observatory Space Weather Simulation Chain}},
  journal = {Space Weather},
  volume = {16},
  number = {5},
  pages = {461-475},
  doi = {10.1002/2017SW001774},
  year = {2018}
}

@article{Sidelnik2023gamma,
title = {{The capability of water Cherenkov detectors arrays of the LAGO project to detect Gamma-Ray Burst and high energy astrophysics sources}},
journal = {Nuclear Instruments and Methods in Physics Research Section A: Accelerators, Spectrometers, Detectors and Associated Equipment},
volume = {1056},
pages = {168576},
year = {2023},
issn = {0168-9002},
doi = {10.1016/j.nima.2023.168576},
author = {I. Sidelnik and L. Otiniano and C. Sarmiento-Cano and J.R. Sacahui and H. Asorey and A.J. Rubio-Montero and R. Mayo-Garcia}
}

@article{sarmiento2022arti,
  title="{The ARTI framework: cosmic rays atmospheric background simulations}",
  author={Sarmiento-Cano, C. and Suárez-Durán, M. and Calderón-Ardila, R. and others},
  journal={Eur. Phys. J. C},
  volume={82},
  number={},
  pages={1019},
  year={2022},
  publisher={Springer},
  collaboration={LAGO},
  doi = {10.1140/epjc/s10052-022-10883-z}
}

@article{taboada2022meiga,
  title={{Meiga, a Dedicated Framework Used for Muography Applications}},
  author={Taboada, A and Sarmiento-Cano, C and Sedoski, A and Asorey, H},
  journal={Journal for Advanced Instrumentation in Science},
  volume={2022},
  pages={266},
  doi = {10.31526/jais.2022.266},
  year={2022}
}

@Article{torres2024,
AUTHOR = {Torres Peralta, Ticiano Jorge and Molina, Maria Graciela and Asorey, Hernan and Sidelnik, Ivan and Rubio-Montero, Antonio Juan and Dasso, Sergio and Mayo-Garcia, Rafael and Taboada, Alvaro and Otiniano, Luis and for the LAGO Collaboration},
title={{Enhanced Particle Classification in Water Cherenkov Detectors Using Machine Learning: Modeling and Validation with Monte Carlo Simulation Datasets}},
JOURNAL = {Atmosphere},
VOLUME = {15},
YEAR = {2024},
NUMBER = {9},
ARTICLE-NUMBER = {1039},
DOI = {10.3390/atmos15091039}
}

@article{santos2023first,
  title={{First measurements of periodicities and anisotropies of cosmic ray flux observed with a water-Cherenkov detector at the Marambio Antarctic base}},
  author={Santos, Noelia Ayelen and Dasso, Sergio and Gulisano, Adriana Maria and Areso, Omar and Pereira, Mat{\'\i}as and Asorey, Hern{\'a}n and Rubinstein, Lucas and LAGO collaboration and others},
  journal={Advances in Space Research},
  volume={71},
  number={6},
  pages={2967--2976},
  year={2023},
  publisher={Elsevier},
  doi = {10.1016/j.asr.2022.11.041}
}

@article{asorey2023acorde,
  title={{ACORDE: A new application for estimating the dose absorbed by passengers and crews in commercial flights}},
  author={Asorey, Hern{\'a}n and Su{\'a}rez-Dur{\'a}n, Mauricio and Mayo-Garc{\'\i}a, Rafael},
  journal={Applied Radiation and Isotopes},
  volume={196},
  pages={110752},
  year={2023},
  publisher={Elsevier},
  doi = {10.1016/j.apradiso.2023.110752}
}

\end{document}